**Density and potential wake past an insulating obstacle in a partially magnetized flowing plasma**

S. Das[1,2,a)] and S.K.Karkari[1,2,b)]

[1]Institute for Plasma Research, Bhat, Gandhinagar, Gujarat 382428, India

[2]Homi Bhabha National Institute, Training School Complex, Anushakti Nagar, Mumbai 40094, India

E-mail: a)satadal.das@ipr.res.in, b)skarkari@ipr.res.in

**Abstract:**

The radial characteristics of plasma potential and density around an insulating disc obstacle, placed inside a partially magnetized plasma flow created in cylindrical chamber by hot cathode filament are presented. In the absence of obstacle, centrally sharp minima in potential and maxima in plasma density is observed; however when a macroscopic obstacle is introduced in plasma flow, a clear radially off-centred minima in plasma potential is observed having plasma density peaking near the edge of the obstacle. The depth of potential around the obstacle depends on the axial magnetic field strength. This off-centred radial potential profile in the plasma flow gives rise to focusing of ions around the obstacle edge. Experimentally it is found that the drift velocity of focused positive ions is directly depended on the magnetic field strength and axial positive ion flow velocity. A phenomenological model based on short-circuiting effect is applied to explain the plasma density and potential in the wake region.

1. **Introduction**

    Recently the potential structure around solid obstacle in low temperature partially magnetized flowing plasma have attracted considerable attention both theoretically and experimentally [1-10]. The potential structure behind the obstacle critically depends on the plasma conditions [11]. An obstacle in a low-temperature ionized gas becomes charged by absorbing electrons and ions from the plasma flow [12]. Charging of a spacecraft in ionosphere is identical of solid object in flowing plasma [13,14]. This was the subject of early theoretical [15] and numerical [16] studies. In downstream of the spacecraft there is a region so called wake region, where ion density reduces drastically. The wake itself is due to finite size of the object, while the ion focusing is an electrical effect that will appear regardless of the object size. In laboratory experiments both the wake and focus regions have been detected for un-magnetized case [17-19]. Often particulates are found in a sheath or double layer. In this case an electric field accelerates the ions to supersonic velocities, as required by the Bohm sheath criterion. By using particle-in-cell simulation, Choi and Kushner [20] showed that wake and focusing is important for the charging of dust particles in

plasma. One interesting feature is the depth of potential region on the size of the solid obstacle as reported in [21]. The interaction of plasma around solid obstacle makes the flow highly nonlinear and creates some oscillatory instabilities. These disturbances in the plasma induces an electric dipole moment on the conducting object. This is already observed for different radii of the object [22,23]. The charge distribution becomes more conspicuous for insulating objects. It is observed that due to the flow, an electric dipole moment will develop, which strongly influences the surrounding plasma [22, 24-26], especially ion trajectory. As ion trajectories are highly distorted due to different plasma transport, it is interesting to see how this effects the ion focusing. Theoretical techniques can be used to find approximate solutions for the electric field and plasma density distribution around the solid object. The magnetic field introduces a remarkable effect on spatial density as well as potential characteristics. This is caused due to significant disparity in the transport rate of electrons and positive ions across the magnetic field. The intrinsic electric fields generate $\vec{E} \times \vec{B}$ drifts, which lead to cross-field turbulent transport [27-29]. Experimentally it is observed that in absence of solid object, axial magnetic field creates a symmetric potential structure to impede the positive ions at the centre [30], but by introducing a solid object at the centre of plasma column, modifies the potential structure into an off-centred one to focus the ions around it.

The wake structure that results from the interaction between flowing plasma and solid object is important in case of spacecraft moving at very high speed in ionosphere. In early 1990's, the ion focusing problem around obstacle motivated the researchers to develop theoretical models and simulations [1,2,31-33]. Previously most of the laboratory experiments on flowing plasma past conducting obstacles have been carried out in the absence of magnetic field or with magnetic field sufficiently strong to magnetize the ions [3-5,8,9]. It is also important to identify the effect of axial magnetic field on ion focusing in presence of insulating object.

In this paper, we present an experimental study of radial plasma density and potential behaviour inside a magnetized flowing plasma in presence of floating insulating solid object. The experimental results indicate an off-centred radial plasma potential and density profile, which reflects ion impediment phenomena inside the cylindrical plasma column. It is also found that the radial plasma density and potential behaves oppositely in presence of axial magnetic field which is explained based on the phenomenological model described in [30].The paper has been organized as follows. The experimental setup is briefly described in section 2. In section 3, the phenomenological model has been presented along with the experimental results. The important results are further discussed in section 4 and the outcome of the work has been summarized in section 5.

## 2. The Experimental Setup:

The experiment is carried in a cylindrical vacuum chamber shown in Fig. 1(a). The chamber is evacuated to a base pressure $2 \times 10^{-5}$ mbar with 375 l$s^{-1}$ Turbo-molecular pump. The axial magnetic field is produced by three electromagnet coils made from copper strep wound in double pan-cake configuration. The coils are assembled in Helmholtz configuration to produce axial magnetic field B = 16 mTesla at the centre for coil current 60A. Fig. 1(b) shows the axial magnetic field plots. The highlighted region shown on the plots corresponds to uniform magnetic field region over a distance of 28 cm.

The plasma is produced in argon with hot tungsten filament fitted in an annular ring, which acts as cathode and a circular anode grid. The tungsten filaments (diameter 0.2 mm and length 12 cm) are heated by passing 18 A – 20 A alternating current at 50 Hz, provided by a step-down transformer. For extracting the electrons from the filament, the centre tap of transformer is biased at -80 V with respect to the grounded chamber. The annular ring having tungsten filaments are placed axially, at z = 15 cm of the linear cylinder. In the entire experiment, the pressure was kept at 0.2 Pa. Plasma parameters have been measured by using cylindrical Langmuir Probe from the radial port as shown in fig-1. The sweep voltage (-65 V to +25 V) to the probe is given by a function generator through an amplifier and the measured data is observed as well as stored in oscilloscope.

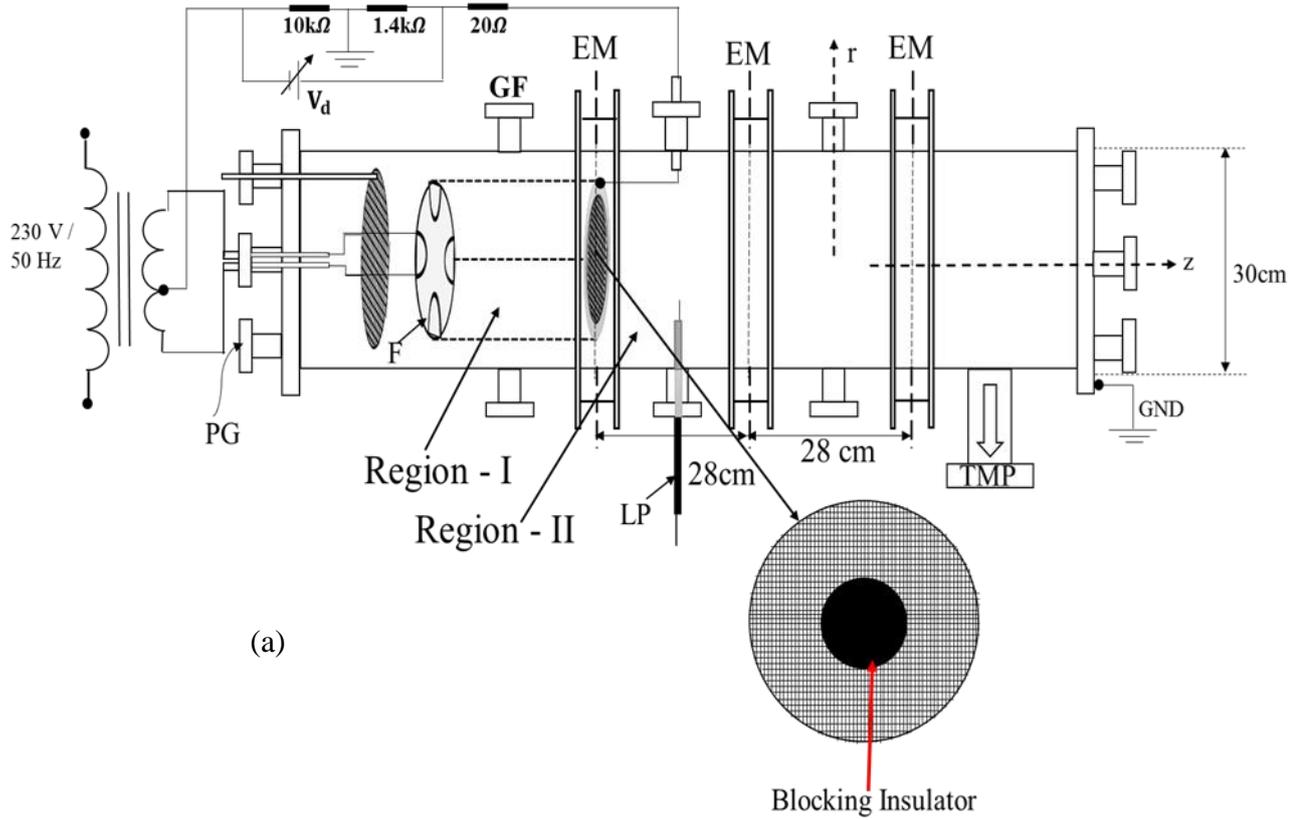

**Figure 1.** (a): Schematic of the experimental setup: EM – electro-magnets, $V_d$ – discharge voltage, PG – pressure gauge, GND – ground, LP – cylindrical Langmuir probe, TMP – Turbo Molecular Pump. (b) Plot of axial magnetic field between the electro-magnet coils; marked region shows the region between the EM coils.

### 3. Phenomenological model and the Experimental results:

In the experiment, the plasma is created by the ionizing electrons from the filament, which are then accelerated by the gridded anode. This separates the plasma region in to two parts Region-1 & Region-II as indicated in Fig. 1(a). The region-II is basically created by the flow of electrons and positive ions that passes through the gridded anode. Experimentally it is observed that region-I always remain at a higher potential than region-II. This potential difference shown in Fig. 2, which is found to be a few volts (~10-16 V) that is much lower than the ionization potential of argon (15.6 eV) results in the flow of positive ions from region-I to region-II. The potential in region – I and region – II is represented as $\varphi_1$ and $\varphi_1$ respectively. The positive ions are charge compensated by the background electrons.

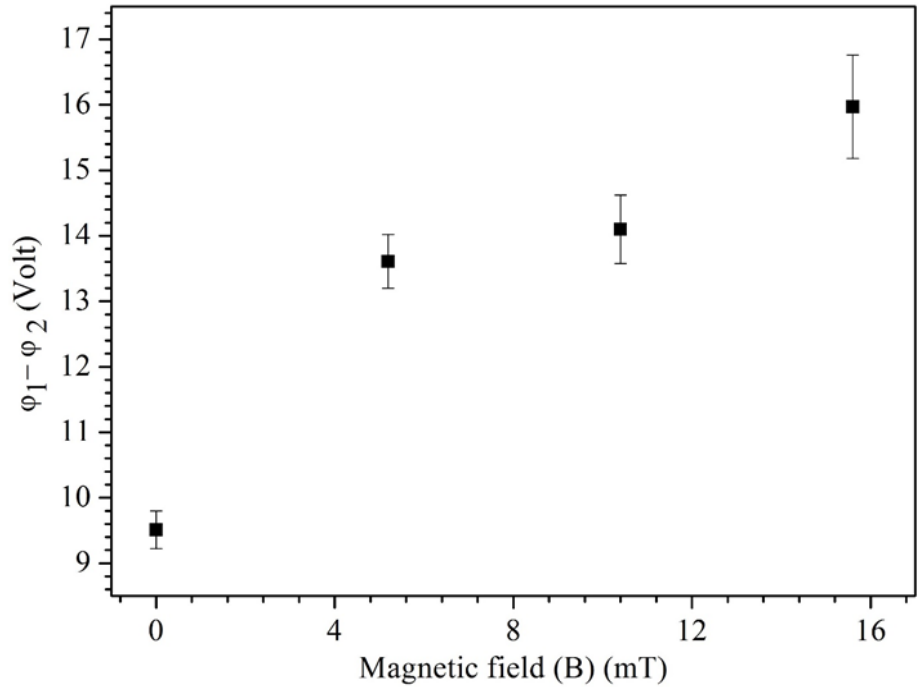

**Figure 2.** Variation of potential difference between Region – I and Region – II with axial magnetic field.

In order to create a wake in the flowing plasma, the central region of the gridded anode is partially shielded with an insulating disc made of a mica sheet of diameter 60 mm and thickness 1.0 mm. The radial density and plasma potential in the wake region at a distance of 15 cm from the obstruction is measured using a single Langmuir probe and the corresponding results are presented in Fig. 4 and 5. In the absence of axial magnetic field, the plasma density shows a small fall in the centre of the discharge and it peaks at the edge of the obstacle. On the other hand, the plasma potential is found to be consistently higher in the centre. When the axial magnetic field is introduced, the central plasma density decreases by almost 50 %- 60%.

In order to model the effect of insulating object on radial plasma properties inside the wake region, we assume a cylindrical plasma tube, having uniform axial magnetic field as schematically shown in Fig. 1(a). The magnetic field intersects a grounded end plate at the far end from the obstacle. This setup resembles the short-circuiting phenomena observed in our previous report [30]. In partially magnetized plasma, the radial motion of electrons is restricted by the axial magnetic field, whereas the positive ions can cross the magnetic field easily. This violates the ambipolar diffusion across the magnetic field lines; however the electrons flow along the magnetic field lines to reach the conducting grounded end plate along with the positive ions and recombine at the wall surface. The difference in the flux of electrons reaching the central region and outside constitutes a virtual current through the conducting plate. In this way the global ambipolarity is maintained in the partially magnetized plasma system. To simplify the above problem, we provide a schematic for cross-section of magnetized plasma column as shown in the Fig. 3.

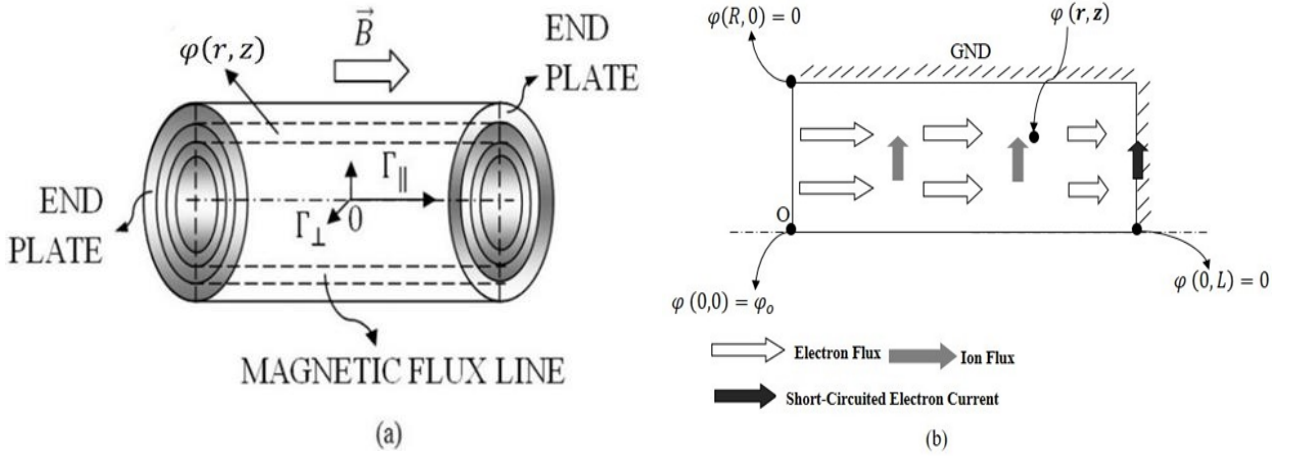

Figure 3. (a) Schematic of magnetized plasma column showing equipotential plasma $\varphi(r,z)$ at any point along the magnetic field flux surfaces; (b) highlighting a rectangular cross-section of the cylindrical column, with magnetic field along the z-axis.

### 3.1 *Radial Plasma Density:*

Since plasma under consideration is in steady state, hence plasma density in the discharge can be calculated by simultaneously solving the flux continuity equations for both positive ions and electrons as shown in Ref.30;

$$\vec{\nabla} \cdot \vec{\Gamma}_i = S, \qquad \vec{\nabla} \cdot \vec{\Gamma}_e = S \tag{3.1}$$

Here S is the ionization source term, whereas $\vec{\Gamma}_i$ and $\vec{\Gamma}_e$ is the flux of positive ions and electrons. $\Gamma_e$ and $\Gamma_{ion}$, can be expanded in cylindrical coordinates as written in Eqs. (3.2) and (3.3) [34].

$$D_{\perp i}\frac{d^2 n_i}{dR^2} + \frac{1}{R}D_{\perp i}\frac{dn_i}{dR} + \frac{D_{\perp i}}{T_i}\frac{d}{dR}\left(n_i \frac{d\phi}{dR}\right) + \frac{D_{\parallel i}}{T_i}\frac{d}{dz}\left(n_i \frac{d\phi}{dz}\right) = -\nu_{iz} n_i \tag{3.2}$$

$$D_{\perp e}\frac{d^2 n_e}{dR^2} + \frac{1}{R}D_{\perp e}\frac{dn_e}{dR} - \frac{D_{\perp e}}{T_e}\frac{d}{dR}\left(n_e \frac{d\phi}{dR}\right) - \frac{D_{\parallel e}}{T_e}\frac{d}{dz}\left(n_e \frac{d\phi}{dz}\right) = -\nu_{iz} n_e \tag{3.3}$$

In the experiment, it is observed that the axial gradient scale length is sufficiently large as compared to the radial length hence z-dependence is basically ignored. The above approximation is also valid for an infinitely long plasma tube. The diffusion loss of charge particles across magnetic field line is compensated by production of electron ion pair by electron impact ionization at a rate $\nu_{iz}$, where $\nu_{iz}$ represents ionization frequency. In the experiment, it is observed that the peak density happens at the edge of the obstacle. Hence we can solve the above equation as a function of R, where R = ( r − $r_1$ ); $r_1$ is the radius of solid object = 3 cm and r is radius of grounded chamber.

Taking in to consideration, the plasma parameters $T_e$ =2.3 to 3.3 eV, $T_i$ = 0.025 eV, $\nu_{iz} \sim 10^3$ Hz; the perpendicular diffusion coefficient for electrons $D_{\perp,e}$ in the present experiments is found to be smaller than the parallel diffusion constant $D_{\parallel,e}$; i.e $D_{\perp,e} \ll D_{\parallel,e}$. Furthermore, the quasi-neutrality condition is valid along the magnetic field, $n_i \cong n_e = n$, hence ignoring the sheath region, the axial electric field $\frac{d\phi}{dz}$ in Eqs (3.2) and (3.3) can be approximated as zero [16]. Based on these basic assumptions, the above equation reduces to;

$$D_{\perp i}D_{\perp e}(T_i + T_e)\frac{d^2n}{dR^2} + D_{\perp i}D_{\perp e}(T_i + T_e)\frac{1}{R}\frac{dn}{dR} = -\nu_{iz}(T_e D_{\perp i} + T_i D_{\perp e})n \qquad (3.4)$$

Furthermore the terms containing $T_i$ and $D_{\perp,e}$ can be ignored provided $T_i \ll T_e$ and $D_{\perp,e} \ll D_{\perp,i}$ which is greatly applicable in this case.

Therefore Eq. (3.4) reduces to a simpler form,

$$\frac{d^2n}{dR^2} + \frac{1}{R}\frac{dn}{dR} + \frac{\nu_{iz}}{D_{\perp,e}}n = 0 \qquad (3.5)$$

Equation (3.5) can be can be solved with appropriate boundary conditions as n(R = 0) = $n_0$ and n(R = 10) = 0, to obtain the radial density inside the plasma column.

At, r = 3.0 cm, R = 0 cm and r = 13.0 cm, R = 10 cm.

By using the above boundary conditions we can express the solution of Eq. (3.5) as,

$$n(R) = n_0 J_0(\gamma R) \qquad (3.6)$$

Where, $J_0(\gamma r)$ is zeroth-order Bessel's function and,

$$\gamma = \left(\frac{\nu_{iz}}{D_{\perp,e}}\right)^{\frac{1}{2}} \qquad (3.7)$$

For pressure = 0.2Pa, $\nu_c = 2.03 \times 10^6$ Hz; $\omega_{ce} = 9.13 \times 10^8$, $1.83 \times 10^9$, $2.74 \times 10^9$ Hz for magnetic field strength B = 5.2, 10.4, 15.6 mT respectively. As $\omega_{ce} \gg \nu_c$ (Condition for magnetized plasma); hence the coefficient $D_{\perp,e} \approx \frac{k_B T_e \nu_c}{m_e \omega_{ce}^2}$; where $\nu_c$ and $\omega_{ce}$ are electron neutral collision frequency and electron cyclotron frequency respectively. By substituting R = ( r − $r_1$ ) in Eq. (3.6) we can express the normalized plasma density as,

$$N(r - r_1) = J_0\{\gamma(r - r_1)\} \tag{3.8}$$

Eq. (3.8) gives the radial density variation in the backside of solid obstacle. The effect of magnetic field is absorbed implicitly in $\gamma$.

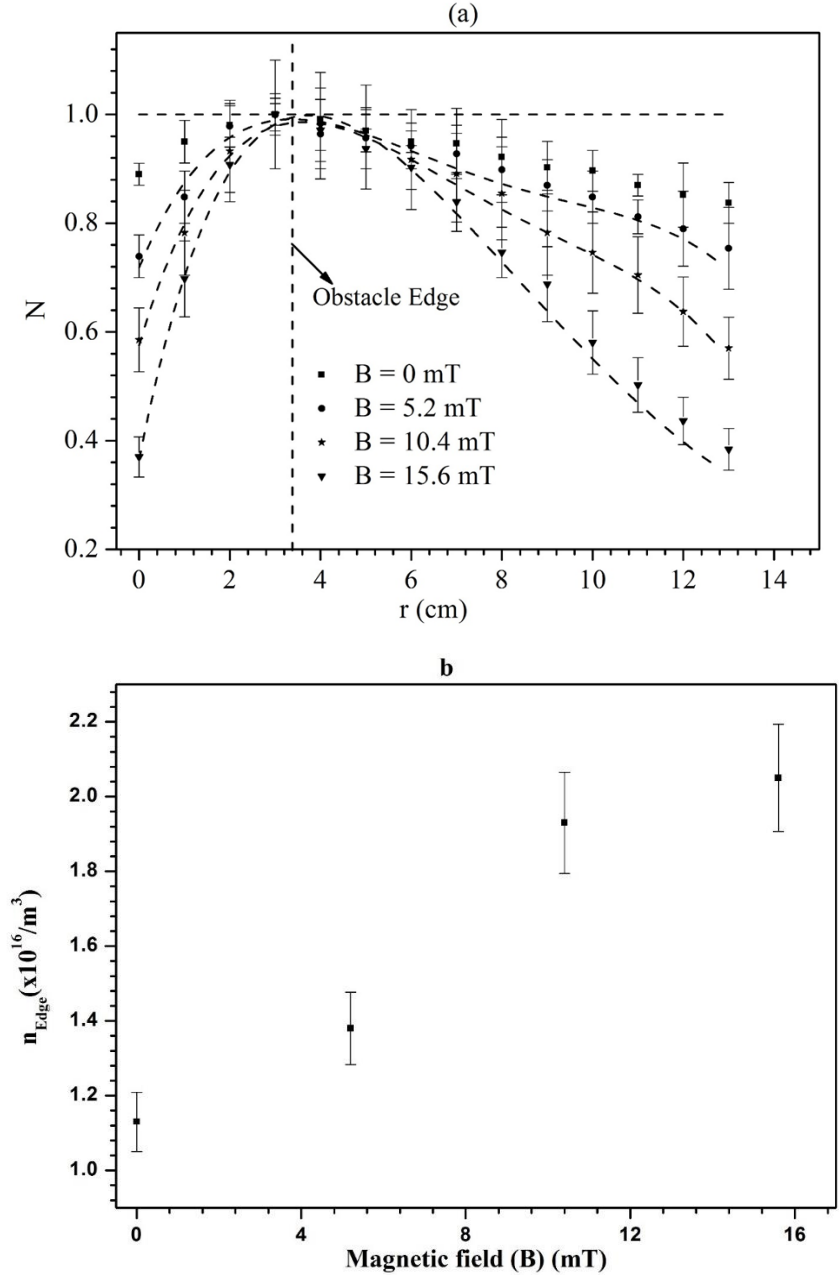

**Figure 4.** Graph of radial plasma density variation for P = 0.2 Pa (a). Dash lines are theoretical plots and scattered points are experimental plots. (b): Graph of Edge density

The calculated radial plasma density using Eq. (3.8) along with experimental data for different axial magnetic field strengths is shown in Fig.4 (a). The edge of the object coincide with the peak plasma density around r = 3.0 cm. As the magnetic field increases, the plasma density falls in centre as well as towards wall.

## 3.2 Radial Plasma Potential:

The radial plasma potential can be calculated by applying the ambipolarity condition, in which the radial motion of positive ions is balanced by the axial flow of electrons towards the grounded end plate. This is the well known Simon short-circuit effect [35]. As discussed in Ref [30], the cross-field electron flux, is equivalent to the short-circuited current as follows.

$$\Gamma_{\perp e} = \Gamma_{short\ circuited,e} \approx \Gamma_{eo} J_0(\gamma R) \exp\left[-\frac{\{\varphi_o(R) - \varphi_o(R=0)\}}{T_e}\right] \tag{3.9}$$

The combined ambipolarity is satisfied when;

$$\Gamma_{short\ circuited,e} = \Gamma_{\perp i} \tag{3.10}$$

On substituting the expressions for different fluxes, $\varphi_0(R)$ as $\varphi(R)$ and $\varphi_o(R=0)$ as $\varphi_o$ in Eq. (3.10), we get;

$$n_0 J_0(\gamma R) \exp\left[\frac{\{\varphi_0 - \varphi(R)\}}{T_e}\right] v_{th} = [\mu_{\perp i} n E_\perp - D_{\perp i} \nabla_\perp n] \tag{3.11}$$

$$v_{th} \exp\left[\frac{e\{\varphi_0 - \varphi(R)\}}{T_e}\right] = \frac{\mu_{\perp i} N}{J_0(\gamma r)} \frac{d\varphi(R)}{dR} + \frac{D_{\perp i}}{J_0(\gamma R)} \frac{dN(R)}{dR} \tag{3.12}$$

The above equation can be further simplified by considering that the un-magnetized positive ions, which are typically at room temperature, giving $\frac{D_{\perp i}}{\mu_{\perp i}} \ll 1$.

Hence Eq. (3.12) which finally reduces to,

$$v_{th} \exp\left[\frac{e\{\varphi_0 - \varphi(R)\}}{T_e}\right] = -\frac{\mu_{\perp i} N \gamma}{J_0(\gamma R)} \frac{J_1(\gamma R)}{dN} \frac{d\varphi(R)}{dN} \tag{3.13}$$

The first derivative of the zeroth order Bessel function, $\frac{d}{dR} J_0(R) = -J_1(R)$; $J_1(R)$ = First order Bessel function.

Therefore final expression for the radial potential profile is as follows:

$$\varphi(R) = \varphi_0 - \frac{T_e}{e}\left[\ln\left(\frac{\mu_{\perp i}\gamma}{v_{th}}\right) + \ln\left(\frac{T_e}{e}\right) + \ln\left\{\frac{J_1(\gamma R)}{J_0(\gamma R)}\right\} - \ln\{-\ln(J_0(\gamma R))\}\right] \tag{3.14}$$

Again substituting $R = (r_1 - r)$ in Eq. (3.14) we can express the normalized plasma density in terms of radial distance with respect to the peak density at r = r1,

$$\varphi(r - r_1) = \varphi_0 - \frac{T_e}{e}\left[\ln\left(\frac{\mu_{\perp i}\gamma}{v_{th}}\right) + \ln\left(\frac{T_e}{e}\right) + \ln\left\{\frac{J_1\{\gamma(r - r_1)\}}{J_0\{\gamma(r - r_1)\}}\right\} - \ln\{-\ln(J_0(\gamma(r - r_1)))\}\right] \tag{3.15}$$

In Fig.5, the radial potential profile are plotted along with the experimental data for different axial magnetic fields.

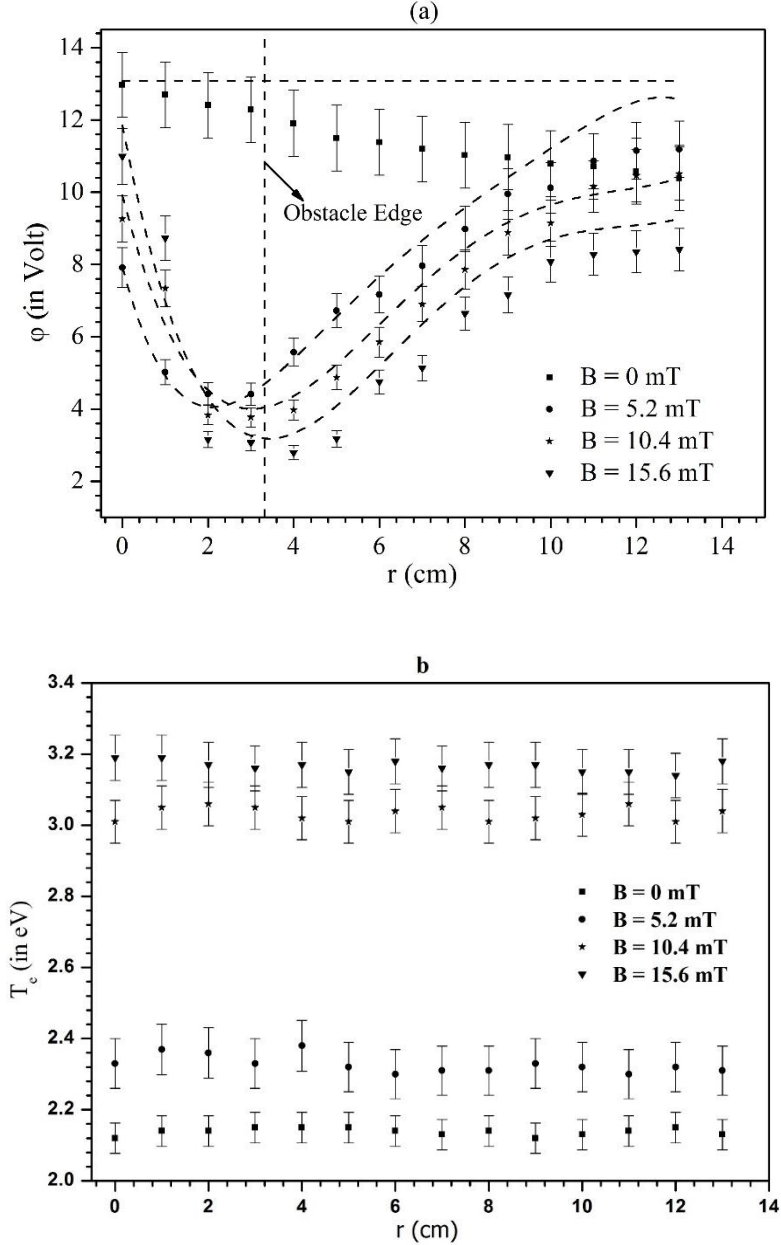

**Figure 5.** Plot of radial plasma potential variation with different magnetic fields for P = 0.2 Pa (a). Dash lines are theoretical plots and scattered points are experimental plots

The experimentally measured electron temperature ($T_e$) plotted in Fig. 5(b), suggests that $T_e$ remains almost constant over the entire region.

4. **Discussion**:

   In the previous papers, the wake potential behind an obstacle placed inside magnetized flowing plasmas had been reported by a few authors [2,10,19,36,37]. In the present experiment, this flow is generated due to the potential difference created between the source region and the drift region. In Fig. 2, the potential difference Δφ varies between 10 V to 16 V by changing the axial magnetic field strengths. This potential difference gives a free fall velocity to the positive ions ($v_{ion}$), given by $\sqrt{2e\Delta\varphi/M}$ as given in Table-I.

Table-I

Variation of Positive ion velocity between Region – I and II with axial magnetic field strength:

| Magnetic Field (B) (in mTesla) | Δφ (in Volts) | $v_{ion}=\sqrt{2e\Delta\varphi/M}$ (in km/s) |
|---|---|---|
| 0 | 9.51 | 1.51 |
| 5.2 | 13.61 | 1.80 |
| 10.4 | 14.10 | 1.84 |
| 15.6 | 15.97 | 1.95 |

The magnitude of $v_{ion}$ comes to be lesser than ion acoustic speed which is 2.6 km/s corresponds to Te = 3.0 eV. In reference [2,19] reported that a significant difference between electron and positive ion speed is the basic requirement for asymmetric screening of potential around an obstacle in a flowing plasma. The subsonic ions on interacting with the edge of the obstacles gets scattered in the shadow region up to a distance which is typically on the order of ion-neutral collision free path (~ 2.0 cm). Therefore the wake region exists for a certain distance behind the obstacle, which again depends on the speed of the positive ions in the flow. It is also observed that as the magnetic field increases, the positive ions gradually gets magnetized. The perpendicular component of ion velocity comes in the picture after the ions gets deflected from the edge of the obstacle. The drift velocity of positive ions in the wake region is assumed to be same as its perpendicular velocity. In contrast the electrons are throughout magnetized for the entire range of magnetic field. The larmor radius of positive ions and electrons are given in Table-II.

Table-II

Variation of larmor radius for Electrons and Positive ion with axial magnetic field strength:

| Magnetic Field (B) (in mTesla) | $r_{L,electron}$ (in cm) | $r_{L,ion}$ (in cm) |
|---|---|---|
| 5.2 | 0.07 | 14.4 |
| 10.4 | 0.03 | 7.40 |
| 15.6 | 0.02 | 5.20 |

As seen in Fig. 5(a), the depth of potential well is found to be significant at maximum magnetic field strength. The potential minima also shifts towards the edge and more sharp at higher magnetic field strength, which indicates the focusing of positive ions are stronger at the edge of obstacle. Fig. 6 shows the variation of ion larmor radius and the distance of potential minima from the edge of the obstacle (d) ratio, as function of axial magnetic field strengths.

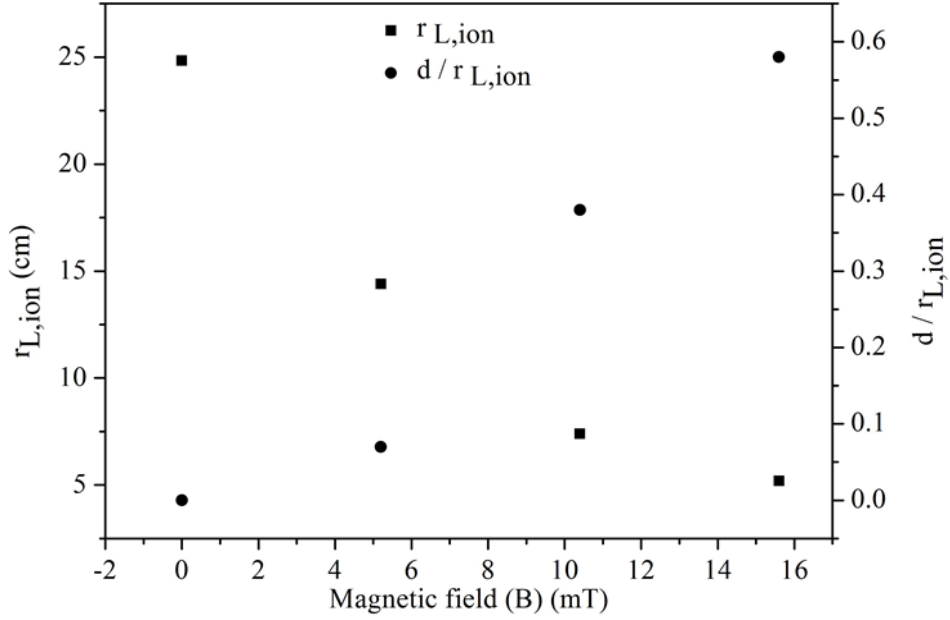

**Figure 6.** Variation of $r_{L,ion}$ and $d/r_{L,ion}$ for different magnetic field strength.

From the above graph (Fig. 6), it is clear that with increasing axial magnetic field the positive ions are becoming magnetized in wake region, which makes them more electrostatically focused at the edge of obstacle. The resultant electric field around the obstacle in the wake region gives rise to less radial deflection of positive ions towards the edge, which creates a focused plasma flow around the obstacle.

Here from Table – I and Fig. 6, it is clear that the position of potential minima is dependent on positive ion flow velocity, which effects the electrostatic focusing of ions.

From Fig. 5 (a), it is observed that the potential around the obstacle becomes more positive with increasing axial magnetic field strength because of two reasons:

1. Positive ions are more focused around the obstacle with increasing axial magnetic field strength, which will create more positive space charge around it.
2. Due to less crossed-field diffusion electrons are not able to cross the magnetic field lines as compare to ions towards the obstacle, which produce positive space charge in front of the obstacle.

Previously the radial plasma density variation in wake region was not reported experimentally. In the present experiment, it is found that the density variation in the wake goes against the potential variation. The plasma density peaks at the edge of obstacle, whereas the potential is minimum there. This leads to an electric field which drives and focuses the positive ions towards its axis. The phenomenological model which is based on short-circuiting effect, can explain the observed variations. It is astonishing that a simple diffusion equation based on particle balance can explain the observed effects and also proves the existence of short-circuiting effect contributing to the preferable density and potential profile inside the setup. The model also predicts the role of magnetic field on the magnitude of potential well created in the wake region. In absence of magnetic field, both electrons and positive ions can diffuse in the centre easily, therefore the height of the potential minima is observed to be small. On the other hand as the magnetic field increases, the cross field motion of electrons towards the obstacle as well as to the side wall of the grounded chamber becomes stringent. As a result, the un-magnetized ions, which have a directed flow due to potential difference in the source and the diffusion regions can move in all the directions in the wake region. This creates such contradicting density and potential behaviour.

5. **Conclusion:**

Summarizing the overall contents of the paper, based on Simon short-circuiting effect the radial plasma potential and density profile in presence of insulating circular object inside a magnetized plasma column has been explained. It has been shown that due to presence of solid insulating object in plasma flow an ion focusing region having higher ion density around it. Un-magnetized positive ion flow creates an asymmetric potential and density profile around the object. It is found that the convergence of ion flow or ion focusing around the solid obstacle is dependent on ion flow velocity. A contradictory behaviour in plasma density and potential has been observed in presence of axial magnetic field and solid insulating object. Our experimental results are found to be in good agreement with the phenomenological model as discussed in the paper.